\documentclass[aps,prl,twocolumn,superscriptaddress]{revtex4-2}
\usepackage{graphicx}
\usepackage{color}
\usepackage{amsmath,amssymb,bm,graphicx,color,gensymb,bbold,appendix,hyperref}
\usepackage{ulem}

\begin{document}

\title{Magnetic excitations in biaxial-strain detwinned $\alpha$-RuCl$_{3}$}

\author{Yi Li}
\affiliation{Center for Advanced Quantum Studies, School of Physics and Astronomy, Beijing Normal University, Beijing 100875, China}
\affiliation{Key Laboratory of Multiscale Spin Physics, Ministry of Education, Beijing Normal University, Beijing 100875, China}

\author{Yanyan Shangguan}
\author{Xinzhe Wang}
\affiliation{National Laboratory of Solid State Microstructures and Department of Physics, Nanjing University, Nanjing 210093, China}

\author{Ruixian Liu}
\author{Chang Liu}
\affiliation{Center for Advanced Quantum Studies, School of Physics and Astronomy, Beijing Normal University, Beijing 100875, China}
\affiliation{Key Laboratory of Multiscale Spin Physics, Ministry of Education, Beijing Normal University, Beijing 100875, China}

\author{Yongqi Han}
\author{Zhaosheng Wang}
\affiliation{Anhui Key Laboratory of Low-Energy Quantum Materials and Devices, High Magnetic Field Laboratory, HFIPS, Chinese Academy of Sciences, Hefei, Anhui 230031, China}

\author{Christian Balz}
\affiliation{ISIS Neutron and Muon Source, Rutherford Appleton Laboratory, Didcot OX11 0QX, UK}
\affiliation{Neutron Scattering Division, Oak Ridge National Laboratory, Oak Ridge, Tennessee 37831, USA}

\author{Ross Stewart}
\affiliation{ISIS Neutron and Muon Source, Rutherford Appleton Laboratory, Didcot OX11 0QX, UK}

\author{Shun-Li Yu}
\author{Jinsheng Wen}
\author{Jian-Xin Li}
\affiliation{National Laboratory of Solid State Microstructures and Department of Physics, Nanjing University, Nanjing 210093, China}

\author{Xingye Lu}
\email{luxy@bnu.edu.cn}
\affiliation{Center for Advanced Quantum Studies, School of Physics and Astronomy, Beijing Normal University, Beijing 100875, China}
\affiliation{Key Laboratory of Multiscale Spin Physics, Ministry of Education, Beijing Normal University, Beijing 100875, China}

\date{\today}

\begin{abstract}

The honeycomb magnet $\alpha$-RuCl$_{3}$ has been a leading candidate for realizing the Kitaev quantum spin liquid (QSL), but its intrinsic spin dynamics have remained obscured by crystal twinning. Here we apply biaxial anisotropic strain to detwin $\alpha$-RuCl$_{3}$ single crystals and directly visualize the intrinsic magnetic excitations using inelastic neutron scattering. 
We discover that the low-energy spin waves emerge from the $M$ points—transverse to the magnetic Bragg peaks—providing direct evidence of anisotropic magnetic interactions in $\alpha$-RuCl$_{3}$.
The intrinsic spin-wave spectrum imposes stringent constraints on the extended Kitaev Hamiltonian, yielding a refined, quantitatively consistent set of exchange couplings for the zigzag ground state and its low-energy dynamics.
Above the magnon band, we uncover broad excitation continuua: while a twofold-symmetric feature near 6 meV at $\Gamma$ is consistent with bimagnon scattering, the dominant spectral weight forms a sixfold-symmetric continuum extending up to $\sim16$ meV that cannot be explained by conventional magnons. This strongly supports the presence of fractionalized excitations—a hallmark of Kitaev QSL physics. Our findings establish biaxial strain as a powerful symmetry-breaking probe to access the intrinsic spin dynamics of Kitaev materials and provide critical benchmarks for refining theoretical models of quantum magnetism in $\alpha$-RuCl$_{3}$.

\end{abstract} 

\maketitle

\begin{figure*} [htbp!]
    \centering
    \includegraphics[width=17cm]{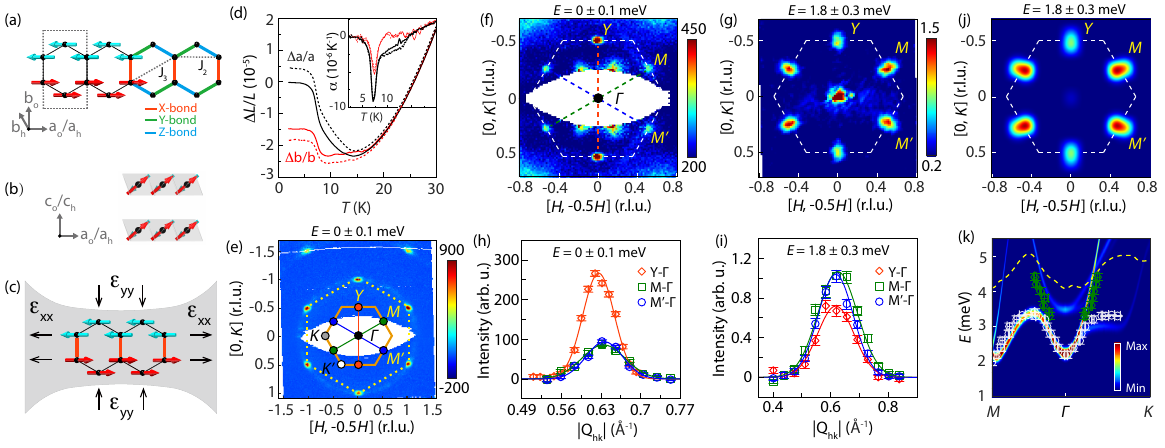}
    \caption{\textbf{Biaxial-strain detwinning, $C_2$ symmetric excitations, and magnetic excitation dispersions.} (a) Schematic for zigzag magnetic order, X-, Y-, Z-bond, and Heisenberg interactions $J_2$ and $J_3$. (b) The layered structure of RuCl$_3$ with spins tilted away from the honeycomb plane by $\alpha\approx31^\circ$. $a, b$ and $c$ mark the lattice axis of the pseudo-orthorhombic (with subscript ${\rm o}$) and $R$\=3 (with subscript ${\rm h}$) structure. (c) A schematic showing that biaxial-anisotropic strain could align shorter Ru-Ru bonds ($b_{\rm o}$ axis), and therefore the zigzag order. (d) Thermal expansion $\Delta L/L$ measured along the $a_{\rm o}$ (black curves) and $b_{\rm o}$ (red curves) axes of a free-standing crystal \cite{SI}. The solid and dashed curves mark different measuring cycles. The inset are corresponding thermal expansion coefficient $\alpha=d(\Delta L/L)/dT$. (e) Nuclear Bragg peaks in $[H, K, 1\pm0.2]$ plane as measured by neutron scattering with $E_i=22$ meV. (f), (g) Magnetic Bragg peaks and low-energy spin waves at the $Y$, $M$, and $M'$ points in the reciprocal space, measured with $E_i=7.4$ meV, with $L=[0.6, 1.2]$ for (f) and $L=[0.6, 3.0]$ for (g). (h), (i) One-dimensional (1D) constant energy cuts along high-symmetry directions $Y-\Gamma$ (red diamonds), $M-\Gamma$ (green squares), and $M'-\Gamma$ (blue circles) extracted from the intensity maps in (f) and (g), with the intensity along the perpendicular directions integrated in the interval $\pm 0.058$ {\AA}$^{-1}$. (j) Constant-energy slice with $E=1.8\pm0.3$ meV calculated from linear-spin-wave theory (LSWT) based on the extended Kitaev model. (k) Energy dispersions (white circles and green diamonds) along $M-\Gamma-K$ extracted from 1D momentum and energy cuts. The colorful intensity maps are single-magnon branches calculated from the LSWT theory. The yellow dashed curves represent the energy mininum of the bimagnon continuum derived from the same calculation. The contribution from the $40\%$ minority twin domains are taken into account in (j) and (k).}
    \label{fig1}
\end{figure*}

The Kitaev model has attracted significant attention as a paradigm for understanding quantum spin liquids (QSLs)—exotic magnetic states characterized by quantum spin entanglement and the absence of conventional magnetic order \cite{Broholm2020, Kitaev2006, Takagi2019, Motome2020review, Matsuda2025}. Defined through bond-dependent anisotropic interactions on a two-dimensional (2D) honeycomb lattice [Fig.~\hyperref[fig1]{1(a)}], the Kitaev model represents one of the few exactly solvable frameworks that predict a QSL ground state. Remarkably, the Kitaev QSL hosts fractionalized excitations, offering a promising route toward realizing fault-tolerant topological quantum computing \cite{Takagi2019}.

Considerable efforts have been dedicated to realizing Kitaev physics in materials and exploring QSL phases with fractionalized excitations \cite{Takagi2019}. Among these materials, $\alpha$-RuCl$_3$ (hereafter RuCl$_3$), a spin-orbit-coupled $J_{\rm eff} = 1/2$ Mott insulator with anisotropic magnetic interactions, has emerged as a leading candidate for realizing Kitaev QSL \cite{Takagi2019,Motome2020review,Plumb2014,johnson15,cao16,Banerjee2017,Do2017,Ran2017,baek2017,Kasahara2018,yokoi2021,bruin2022,Winter2017,Winter2017NC,Kim2024,Kim2022,Park2024,Sears2020,Braden2024}.
RuCl$_3$ is a layered van der Waals material composed of stacked honeycomb lattices of edge-sharing RuCl$_6$ octahedra [Fig. \hyperref[fig1]{1(b)}]. Upon cooling, RuCl$_3$ undergoes a first-order structural transition from $C$2/m to $R$\=3 structure at $T_s\approx150$ K, followed by a magnetic ordering into a zigzag configuration at  $T_N \approx 7$ K  with trilayer (ABC) stacking or $T_N \approx 14$ K  with bilayer (ABAB) stacking 
\cite{Kim2024}.
Below $T_N$, the ordered magnetic moment tilts out of the honeycomb plane by $\alpha\approx31^\circ$ [Fig. \hyperref[fig1]{1(b)}] \cite{johnson15,cao16,Park2024,Sears2020,Kim2024re}. 

Inelastic neutron scattering (INS) studies have provided critical insights into the magnetic excitations of RuCl$_3$. These investigations can be framed within the 2D extended Kitaev model \cite{Rau2014,rethink2020}:
\begin{align}
H = \sum_{ij \in \alpha\beta(\gamma)} \Big[
J \mathbf{S}_i \cdot \mathbf{S}_j +K S^\gamma_i S^\gamma_j 
+\mathit{\Gamma} \left( S^\alpha_i S^\beta_j +S^\beta_i S^\alpha_j\right)\nonumber
\\
\
 +\mathit{\Gamma'} \left( S^\gamma_i S^\alpha_j+S^\gamma_i S^\beta_j+S^\alpha_i S^\gamma_j
+S^\beta_i S^\gamma_j\right)\Big],
\label{eq1}
\end{align}
where $J$, $K$, $\mathit{\Gamma}$, and $\mathit{\Gamma'}$ represent the Heisenberg, Kitaev, the symmetric off-diagonal exchange interactions, respectively. Spin-wave analysis of the low-energy magnons, performed by some of us, has highlighted a dominant ferromagnetic $K$ and a significant $\mathit{\Gamma}$ \cite{Ran2017,Ran2022}.

Subsequent INS experiments uncovered a prominent magnetic excitation continuum spanning $E\approx2-15$ meV at the $\Gamma$ point, which was attributed to fractionalized excitations — a key signature of the Kitaev QSL \cite{Banerjee2017,Do2017}.
However, an alternative interpretation proposed by Winter {\it et al.} suggested that the continuum could arise from incoherent magnetic excitations due to magnon breakdown, driven by strong anisotropic interactions \cite{Winter2017NC, Winter2018PRL}, challenging the attribution to fractionalized spin excitations. Nevertheless, this scenario underestimates the spin-wave energy minimum and remains unverified experimentally.
Moreover, recent INS studies have observed low-energy, magnon-like dispersions at the Brillouin zone center $\Gamma$ \cite{Ran2022,Banerjee2018,Samarakoon2022} that could not be satisfactorily described in previous works \cite{rethink2020, Matsuda2025,Ran2017,Wu2018, Laurell2020, Cookmeyer2018, Sahasrabudhe2020, Suzuki2018,Winter2017NC,Banerjee2018,Samarakoon2022,Ran2022,Winter2018PRL,SI}. 
Furthermore, previous INS studies of RuCl$_3$ were conducted exclusively on twinned samples containing twin domains aligned along three equivalent directions, leading to averaged magnetic excitations with apparent $C_6$ symmetry \cite{cao16,Banerjee2017}. This twinning obscures key aspects of the intrinsic spin dynamics, including potential anisotropies and directional dependencies that are crucial for understanding the underlying magnetic interactions.

These unresolved issues underscore the ongoing challenge of obtaining intrinsic magnetic excitations and developing a minimal theoretical framework that accurately captures the magnetic ground state of RuCl$_3$ — a prerequisite for understanding the emergence of the putative Kitaev QSL under applied magnetic fields \cite{Matsuda2025}.

In this work, we introduce a biaxial anisotropic-strain technique [Fig. \hyperref[fig1]{(c)}] that effectively detwins RuCl$_3$ single crystals, aligning approximately $\sim60\%$ of the magnetic domains along a single in-plane direction [Fig. \hyperref[fig1]{(f),(h)}]. Using inelastic neutron scattering (INS), we directly probed the intrinsic spin dynamics in the detwinned state. The low-energy spin-wave spectrum exhibits twofold ($C_2$) symmetry and reveals an exotic feature: the low-energy magnons emerge primarily from the $M$ and $M'$ points, rather than from the magnetic Bragg positions at the $Y$ points [Figs. \hyperref[fig1]{1(f)–(i)}]. This unconventional behavior provides direct evidence for strongly bond-directional anisotropic magnetic interactions in RuCl$_3$, which are otherwise masked in twinned crystals.

The observed twofold-symmetric magnon dispersion of the zigzag-ordered state can be captured by the two-dimensional extended Kitaev model [Eq. \hyperref[eq1]{(1)}] using the following exchange parameters: $J = -1.47$, $K = -11$, $\Gamma = 3.52$, $\Gamma' = 0.33$, $J_2 = -0.91$, and $J_3 = 1.89$ meV [Figs. \hyperref[fig1]{(j),(k)}]. In addition to reaffirming the dominant ferromagnetic $K$ and sizable off-diagonal $\Gamma$ couplings in RuCl$_3$ \cite{Ran2017,Wang2017PRB}, this refined parameter set offers a more accurate microscopic basis for understanding its deviation from the Kitaev quantum spin liquid limit \cite{SI}. With these parameters, the intrinsic magnon spectral weight is significantly suppressed at $Y$, and the residual signal observed there [Fig. \hyperref[fig1]{1(g)}] arise from the contribution from the $\sim40\%$ minority twin domains, whose $M/M'$ points overlap with $Y$ in the twinned geometry \cite{SI}.

This Hamiltonian also predicts a bimagnon band minimum at $E_2^{\rm min} \approx 4.5\pm0.5$ meV (yellow dashed curves in Fig. \hyperref[fig1]{1(k)}). Below this threshold, we observe a $C_6$-symmetric continuum between the single-magnon branches ($E \lesssim 3.3$ meV) and $E_2^{\rm min}$, which cannot be explained by conventional spin-wave theory. Above $E_2^{\rm min}$, a broad peak centered at $\sim6$ meV emerges near $\Gamma$ and displays a twofold intensity profile consistent with the $C_2$ symmetry of the bimagnon density of states. However, the total bimagnon spectral weight accounts for only a small portion of the full response, as the excitation continuum extends to $E \approx 16$ meV. The coexistence of sixfold-symmetric features at $M$, $M'$, and $Y$ and the intense, anisotropic continuum at $\Gamma$—with spectral weight far exceeding that of bimagnons—points to the presence of fractionalized spin excitations in RuCl$_3$.



\hspace*{\fill}

\noindent
\textbf{Biaxial-strain detwinning of RuCl$_3$}

\noindent
Early neutron and x-ray diffraction studies suggested that in the high-temperature $C$2/m phase, one of the three Ru-Ru bonds is approximately $0.2\%$ shorter, indicating an in-plane symmetry-breaking distortion \cite{cao16}. However, more recent high-resolution x-ray diffraction measurements on RuCl$_3$ identify an $R$\=3 structure below $T_s \approx 150$ K and even across the magnetic transition at $T_N$, suggesting equivalent Ru-Ru bond lengths at low temperatures \cite{Kim2024, Park2024}. 
Despite this, the magnetoelastic coupling associated with the $C_2$-symmetric zigzag magnetic order below $T_N$ may induce subtle symmetry breaking from the nominal $R$\=3 symmetry and enable detwinning through biaxial anisotropic strain. 

To verify this, we conducted high-precesion thermal-expansion measurements of RuCl$_3$ single crystals along the orthorhombic $a_{\rm o}$ and $b_{\rm o}$ axes (see the Supplemental Material for details) \cite{SI}. As presented in Fig. \hyperref[fig1]{1(d)}, the thermal expansions along $a_{\rm o}$ ($\Delta a/a$) and $b_{\rm o}$ ($\Delta b/b$) remain identical above approximately $20$ K, consistent with the preservation of the $R\bar{3}$ symmetry. Below $T\sim15$~K, however, the thermal expansions along these two directions begin to deviate from one another, displaying a distinct anisotropy across the magnetic transition at $T_{N}\approx7$ K, with a measurable difference $(\Delta L/L)_a-(\Delta L/L)_b\sim0.002\%$.
This in-plane structural anisotropy likely arises from a subtle (relative) shortening of one pair of Ru–Ru bonds (parallel to the $b_{\rm o}$ axis) below $T_{N}$ [red bonds in Fig. \hyperref[fig1]{1(c)}]. The slight variations in $\Delta L/L$ observed between different measurement cycles may reflect changes in twin-domain populations. 
Although significantly smaller than the $0.1\%-0.5\%$ orthorhombic distortions commonly observed in iron-based superconductors—which readily enable domain alignment under uniaxial strain \cite{Lu2016,liu2025spin}—this subtle yet unambiguous anisotropy suggests a weak but crucial in-plane symmetry breaking, offering a practical route for detwinning RuCl$_3$.

We applied biaxial anisotropic strain—compressive along one Ru-Ru bond direction and tensile along the perpendicular axis—using a custom device that exploits differential thermal expansion between an invar-alloy frame and aluminum sheets [Fig. \hyperref[fig1]{1(c)} and Fig. S2] \cite{SI,liu2025spin}. Neutron diffraction measurements with $E_i = 22$ meV reveal sharp, $C_6$-symmetric nuclear Bragg peaks, indicating precise crystal alignment and the preservation of the $R$\=3 structure [Fig. \hyperref[fig1]{1(e)}]. Subsequent measurements with lower-energy neutrons ($E_i = 7.4$ meV) show that the strain efficiently aligns the zigzag magnetic domains, as evidenced by significantly enhanced magnetic Bragg peaks along the compressive strain direction $Y-\Gamma$ and a corresponding suppression of peaks at $M$ and $M'$ in Figs. \hyperref[fig1]{1(f)} and \hyperref[fig1]{1(h)}. These results indicate that RuCl$_3$ can be successfully detwinned under biaxial anisotropic strain below $T_N$. Meanwhile, we noticed that external strain introduces stacking faults and defects in the samples, resulting in the suppression of the $T_{N}\approx7.5$ K phase and slight enhancement of the $T_{N}\approx10-14$ K phases (see the supplemental material for details) \cite{SI}.

While the applied biaxial strain ($\varepsilon_{xx}-\varepsilon_{yy} \lesssim$ 0.4\%) is sufficient to detwin the zigzag domains, it is an order of magnitude smaller than the $\sim$2–4\% uniaxial strain predicted to measurably renormalize the exchange couplings \cite{kaib2021,yadav2018,vatansever2019}. Consistent with a domain-selection role rather than Hamiltonian tuning, the detwinned crystal exhibits magnon dispersions and bandwidths along $M$–$\Gamma$–$M$ that are the same as those of unstrained, twinned samples \cite{Samarakoon2022,Nakajima2022}. We therefore attribute the symmetry contrasts reported below to single-domain access, not to strain-induced modification of $J$, $K$, or $\Gamma$.

\hspace*{\fill}

\noindent
\textbf{Dichotomy between zigzag order and magnons}

\noindent
In Fig. \hyperref[fig1]{1(h)}, the integrated intensity of the magnetic Bragg peaks at $Y$, $M$, and $M'$ follows the ratio $I_Y$:$I_M$:$I_{M'} = 3$:$1$:$1$, indicating that the sample is partially detwinned, with approximately $60\%$ of the magnetic domains aligned along one direction corresponding to the magnetic Bragg peak at $Y$ \cite{SI}. In this detwinned sample, constant-energy slices of the dynamic structure factor $S(\mathbf{Q}, E)$ with $E = 1.8\pm0.3$ meV reveal stronger magnetic excitations at the $M/M'$ points, transverse to the magnetic wavevector positions ($Y$ points) [Fig. \hyperref[fig1]{1(g)}], where the excitations are approximately $33\%$ weaker [$S_Y$:$S_M$:$S_{M'} = 2$:$3$:$3$ in Fig. \hyperref[fig1]{1(i)}]. 
This behavior stands in stark contrast to conventional magnets, where spin waves typically originate from magnetic wavevectors. 
Such a phenomenon arises as a profound consequence of anisotropic magnetic interactions in the extended Kitaev model, which shifts low-energy magnons from the $Y$ points to the $M/M'$ points \cite{Chun2015, Wang2017PRB, Winter2017NC, Winter2018PRL}. Our findings provide a direct experimental confirmation of the unusual dichotomy between magnetic order and low-energy spin waves in RuCl$_3$, offering direct evidence of significant anisotropic interactions in this system.
Moreover, this dichotomy, revealed in biaxial-strain-detwinned samples, provides a practical diagnostic of intrinsic anisotropic magnetic interactions in putative Kitaev materials.

\begin{figure} [htbp!]
    \centering
    \includegraphics[width=8.5cm]{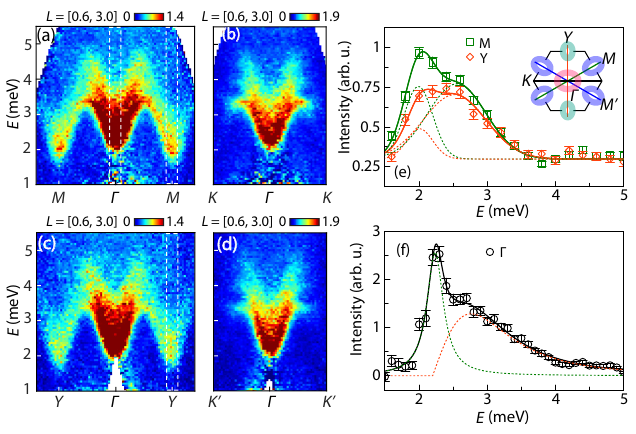}
    \caption{{\bf Energy dependence of magnetic excitations in detwinned RuCl$_3$ below $T_N$.}(a)-(d) Energy-momentum slices of the dynamic structure factor $S(\mathbf{Q}, E)$ along high-symmetry directions: (a), $M-\Gamma-M$, (b), $K-\Gamma-K$, (c) $Y-\Gamma-Y$, and (d) $K'-\Gamma-K'$, measured with $E_i=7.4$ meV. Scattering intensities are integrated along the out-of-plane wavevector $L$ over the range $0.6 \leq L \leq 3$, and along the perpendicular in-plane momentum $Q_\perp$ over the range $Q_\perp = \pm 0.105$ {\AA}$^{-1}$. (e), (f), Energy-dependent cuts at high-symmetry points: $M$ (green squares), $Y$ (red diamonds), and $\Gamma$ (black circles). The momentum interval integrated around these points is $\pm 0.061$ {\AA}$^{-1}$ as marked by dashed rectangles in (a), (c) and Fig. \hyperref[fig3]{3(d)}. The solid curves following the data points are fittings of the data. The error bars represent one standard deviation of the scattering intensity.}
    \label{fig2}
\end{figure}

\begin{figure*} [ht]
    \centering
    \includegraphics[width=17cm]{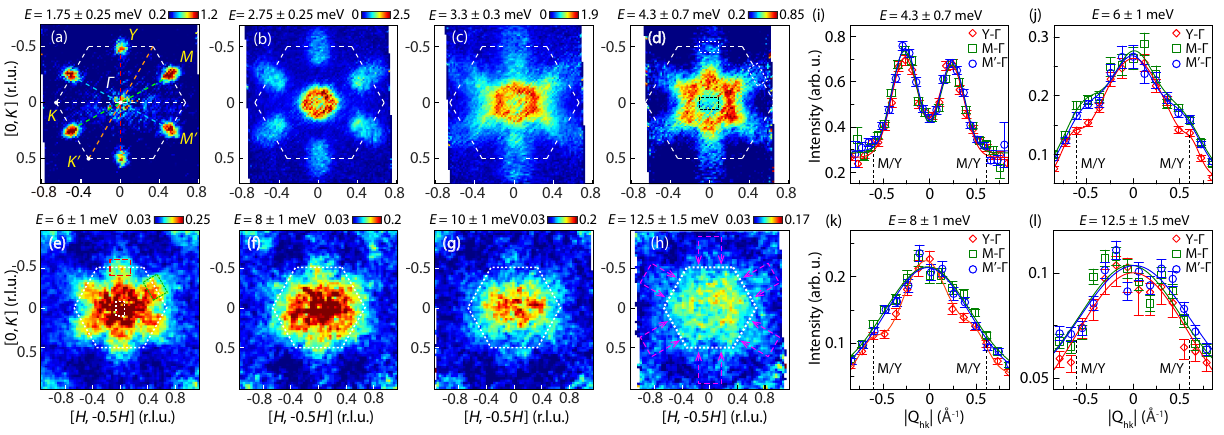}
    \caption{{\bf Wavevector and energy dependence of magnetic excitations in detwinned RuCl$_3$ below $T_N$.} (a)-(h) Constant-energy intensity maps in the $[H, K]$ plane for $E = 1.75\pm0.25$ meV (a), $2.75\pm0.25$ meV (b), $3.3\pm0.3$ meV (c), $4.3\pm0.7$ meV (d), $6.0\pm1.0$ meV (e), $8.0\pm1.0$ meV (f), $10\pm1.0$ meV (g), and $12.5\pm1.5$ meV (h). The white dashed hexagon marks the first Brillouin zone. 
    (i)-(l) Constant-energy moementum cuts along $Y-\Gamma$ (red diamonds), $M-\Gamma$ (green squares), and $M'-\Gamma$ (blue circles) directions extracted from (d), (e), (f), and (h), respectively.
    (a)-(d) and (i) are collected with $E_i=7.4$ meV, while the others with $E_i=22$ meV. Scattering intensities are integrated along the out-of-plane wavevector $L$ over the range $0.6 \leq L \leq 3$ for the slices collected with $E_i=7.4$ meV and $0.6 \leq L \leq 5$ for those measured with $E_i=22$ meV. The solid curves in (i)-(l) are multi-Gaussian fittings of the data points. For the momentum cuts along high symmetry directions in (i)-(l), the integrated momentum interval along the perpendicular directions are $\pm 0.058$ {\AA}$^{-1}$ for (i), and $\pm 0.172$ {\AA}$^{-1}$ for (j)-(l). The error bars represent one standard deviation of the scattering intensity. The vertical dashed lines in (i)-(l) mark the $M/Y$ positions.}
    \label{fig3}
\end{figure*}

\hspace*{\fill}

\noindent
\textbf{$C_2$ and $C_6$ symmetric excitations}

\noindent
Excitations tied to the zigzag order—both spin waves and the bimagnon continuum—inherit the $C_2$ symmetry of the ordered state \cite{Winter2017NC}. By contrast, the fractionalized continuum expected for a proximate Kitaev QSL is Kitaev-dominated and preserves the hexagonal ($C_6$) symmetry of the honeycomb lattice, yielding broad weight near $\Gamma$ and at the zone edges ($M/M'/Y$) \cite{Knolle2014}. Given that our detwinning strain is far below the level required to renormalize the exchange couplings, the observed $C_2$ versus near-$C_6$ patterns provide a symmetry-based diagnostic of the excitation character in RuCl$_3$.

Figures \hyperref[fig2]{2} and \hyperref[fig3]{3} present the magnetic excitations in the detwinned RuCl$_3$ sample measured at $T=2$ K ($<T_N$) with $E_{i}=7.4$ meV and $22$ meV. Figures \hyperref[fig2]{2(a)-(d)} show projections of $S(\mathbf{Q}, E)$ onto the $E-\mathbf{Q}$ planes along the $M-\Gamma-M$, $K-\Gamma-K$, $Y-\Gamma-Y$, and $K'-\Gamma-K'$ directions. These slices reveal dispersive excitations around the $Y$, $M/M'$, and $\Gamma$ points. 
The spin waves emerging from $M$ points are consistent with previous INS studies on twinned samples [Fig. \hyperref[fig2]{2(a)}] \cite{Ran2017, Banerjee2018}, exhibiting a pronounced low-energy magnon peak at $E\approx2$ meV, as shown in Fig. \hyperref[fig2]{2(e)} (green squares).
As noted above, the spectral weight at $Y$ largely reflects contributions from the $M$ and $M'$ points of minority twin domains; accordingly, the band minimum at $Y$ occurs at the same energy but with markedly reduced spectral weight [Figs. \hyperref[fig2]{2(a),(c),(e)}], confirming that the low-energy magnons originate at $M$ rather than at the zigzag wavevector $Y$.
In contrast, the response just above the magnon band ($E\approx2.6$ meV) shows comparable intensity at $M$ and $Y$ [Fig. \hyperref[fig2]{2(e)}] and, in momentum maps, forms an apparently $C_6$-symmetric pattern [Figs. \hyperref[fig3]{3(b)–3(d)}].

At the $\Gamma$ point, a well-defined spin-wave branch forms the lower boundary of the magnetic excitation spectrum and coexists with higher-energy dispersive modes and a broad continuum. As shown in Fig. \hyperref[fig2]{2(f)}, a Lorentzian peak centered at $E\approx2.2$ meV confirms the presence of a coherent low-energy magnon at $\Gamma$, while an excitation continuum extending up to $E\approx4$ meV cannot be attributed to conventional magnons.

The $\Gamma$-point spin waves disperse upward to a maximum energy of $E \approx 3.3$ meV, smoothly connecting to the low-energy magnons emerging from the $M$ points [Fig. \hyperref[fig2]{2(a)}]. Importantly, these $\Gamma$-centered modes carry substantially more spectral weight than those at $M/M'$, highlighting a pronounced ferromagnetic character at $\Gamma$—a feature not adequately captured in previous modeling efforts \cite{Suzuki2021, Samarakoon2022, Matsuda2025}.

The excitation continuum spanning the energy range $E=2-15$ meV at $\Gamma$ reported in privious INS studies were attributed to fractionalized spin excitations \cite{Banerjee2017,Do2017}.
However, an alternative interpretation suggests that this continuum arises from incoherent excitations due to single-magnon decay into bimagnons \cite{Winter2017NC}. In that model, the magnon gap at $M$ was estimated to be $E \approx 0.76$ meV, with a bimagnon minimum of $E^{\rm min}_2 \approx 1.5$ meV, leading to an overlap between the bimagnon continuum and single-magnon bands at both $Y$ and $\Gamma$. In contrast, Fig. \hyperref[fig2]{2} reveals a significantly larger magnon gap at $M$ ($E \approx 2$ meV) and our LSWT calculation generates a bimagnon minimum of $E^{\rm min}_2 \approx 4.5\pm0.5$ meV, indicating that excitations below $\sim 4.5\pm0.5$ meV remain unaffected by magnon breakdown effects. Consequently, the sixfold-symmetric continuum at $M/M'$ and $Y$, and the continuum at $\Gamma$—lying above the single-magnon branch and below $E_2^{\rm min}$—are consistent with fractionalized spin excitations expected in proximity to the Kitaev quantum spin-liquid regime \cite{Ran2022,Knolle2014}.

In Fig. \hyperref[fig3]{3}, constant-energy slices of the magnetic spectrum illustrate how the excitation symmetry evolves with energy in detwinned RuCl$_{3}$. 
At low energies [Fig. \hyperref[fig3]{3}(a)-(c)], we observe well-defined magnons emerging from the high-symmetry $M$ and $M'$ points that clearly exhibit a twofold rotational anisotropy – a direct manifestation of the zigzag order which breaks the sixfold symmetry of the honeycomb lattice. 
Notably, as mentioned above, a diffuse continuum appears around the $M/M'$ and $Y$ points with an almost sixfold symmetric intensity pattern \cite{SI}. 
As the energy increases above the single-magnon band top ($\sim3.3$ meV) [$E=4.3\pm0.7$ meV in Fig. \hyperref[fig3]{3(d)}], the magnetic scattering unexpectedly form a $C_6$ symmetric pattern: the intensity profiles along the $M-\Gamma-M$ and $Y-\Gamma-Y$ directions coincide almost perfectly [Fig. \hyperref[fig3]{3(i)}], signaling an apparent restoration of $C_{6}$ symmetry in the excitation spectrum. 
At still higher energies, in the range above the bimagnon minimum $E_2^{\rm min}$ [Fig. \hyperref[fig3]{3(e)-(g)}], the broad continuum centered at the $\Gamma$ point initially retains a discernible $C_2$ anisotropy (elongated along $K-\Gamma-K$ direction), which is clearly resolved in the comparison of the 1D cuts along high symmetry directions in Fig. \hyperref[fig3]{3(j),(k)}.
However, this anisotropic character progressively weakens with increasing energy, and by the highest energies measured [Fig. \hyperref[fig3]{3(h)}] the excitations form an essentially featureless, isotropic cloud around $\Gamma$ [Fig. \hyperref[fig3]{3(h),(l)}]. 

These results suggest a natural interpretation in terms of coexisting magnon and fractionalized spin excitations. The low-energy $C_2$-symmetric modes are readily identified as single-magnon spin waves of the zigzag order. Below $E_{2}^{\rm min}$, the $C_{6}$-symmetric continuua could be attributed to fractional spin excitations.
With increasing energy, the magnetic response broadens and develops continuum character: in the intermediate $6-10$ meV range, the persistence of $C_2$ anisotropy around $\Gamma$ hints that multi-magnon (e.g. two-magnon) processes tied to the zigzag order are contributing to the spectrum \cite{Winter2017NC}. At yet higher energies, however, the influence of the zigzag order diminishes – the nearly $C_6$-symmetric, diffuse excitations dominating the spectra are inconsistent with conventional magnons and instead point to fractionalized spin excitations arising from the Kitaev interactions in the material \cite{Ran2022}.

\begin{figure} [ht]
    \centering
    \includegraphics[width=8.5cm]{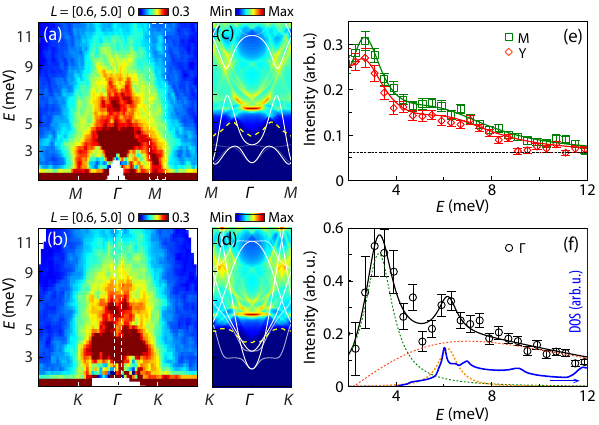}
    \caption{{\bf Excitation continuum and bimagnon intensity.} (a), (b) Energy-momentum slices of the magnetic excitations along $M-\Gamma-M$ (a), and $K-\Gamma-K$ (b). (c), (d), Spin wave dispersion (white solid curves) and two-magnon density of states (color map) calculated from LSWT along $M-\Gamma-M$ and $K-\Gamma-K$.
    (e), (f), Energy-dependent cuts at $M$ (green squares), $Y$ (red diamonds), and $\Gamma$ (black circles) points.  
    The momentum intervals integrated around these points are $\pm 0.122$ {\AA}$^{-1}$ for (e) and $\pm 0.061$ {\AA}$^{-1}$ for (f), as marked by white dashed rectangles in (a) and (b). The blue curve in (f) represents energy-dependent density of states (DOS) for bimagnon continuum at $\Gamma$ integrated within the same momentum interval. The solid curves following the data points are fittings of the data. The error bars represent one standard deviation of the scattering intensity.}
    \label{fig4}
\end{figure}

\hspace*{\fill}

\noindent
\textbf{Magnetic interactions}

\noindent
To determine the magnetic interactions and describe the magnetic excitations, we fit the spin-wave energy dispersions extracted from the 2D slices and 1D cuts [Figs. \hyperref[fig2]{2} and \hyperref[fig3]{3}] using the 2D generic extended Kitaev model [Eq. \hyperref[eq1]{(1)}]. The energy dispersions along high-symmetry directions are shown in Fig. \hyperref[fig1]{1(k)} and Fig. S4(j) (white circles and green diamonds). The best fit yields $J = -1.47$, $K = -11$, $\mathit{\Gamma} = 3.52$, $\mathit{\Gamma'} = 0.33$, $J_2 = -0.91$ and $J_3 = 1.89$ meV, successfully capturing the observed features of the spin waves in partially detwinned RuCl$_3$ [Fig. \hyperref[fig1]{1(j)-(k)}] \cite{SI}. Compared with previous reports \cite{rethink2020, Matsuda2025}, our results highlight the crucial role of minor non-Kitaev, Heisenberg interactions: ferromagnetic $J$, $J_{2}$ and antiferromagnetic $J_3$ in describing the spin waves, in addition to the dominant ferromagnetic Kitaev interaction and a substantial symmetric off-diagonal interaction $\mathit{\Gamma}$ \cite{Ran2017}.

The LSWT calculation for untwinned RuCl$_3$, based on the aforementioned magnetic interactions, predicts significantly different spin-wave dispersions along the $\Gamma-Y$ and $\Gamma-M$ directions [Fig. S4 in the supplemental material] \cite{SI}. While the branches at $Y$ and $M$ exhibit similar energy minima, the excitation at $Y$ carries negligibly weak spectral weight compared to that at $M$, further confirming that anisotropic magnetic interactions shift the low-energy magnetic excitations away from the expected magnetic wavevector ($Y$) to the $M/M'$ points. In our partially detwinned RuCl$_3$ sample, the spectral weight around $Y$ is primarily attributed to spin waves from the $M/M'$ points of the $40\%$ minority twin domains. Taking the partial detwinning effect into account, the calculated results accurately reproduce the observed $C_2$-symmetric constant-energy excitations, as shown in Figs. \hyperref[fig1]{1} and S4.

\hspace*{\fill}

\noindent
\textbf{The $C_2$ symmetric excitation above $E_{2}^{\rm min}$}

\noindent
Figure \hyperref[fig4]{4} presents the magnetic excitations over a broader energy range ($E \approx 2-16$ meV) measured with $E_{i} = 22$ meV. 
The excitation continuum around $\Gamma$ is widespread in reciprocal space [Figs. \hyperref[fig3]{3} and \hyperref[fig4]{4(a),(b)}], leading to broad continua in the energy cuts at the $M$, $Y$, and $\Gamma$ points [Fig. \hyperref[fig4]{4(e),(f)}]. Notably, a prominent peak appears around $E \approx 6\pm1$ meV at $\Gamma$ [orange dashed curve in Fig. \hyperref[fig4]{4(f)}], previously attributed to a spin wave in earlier studies \cite{Banerjee2016}. By comparison, our analysis suggests that the continuua consist of coexisting bimagnons and fractionalized spin excitations. 

To investigate the origin of the excitations, we calculated the density of states (DOS) for bimagnons. Figures \hyperref[fig4]{4(c)} and \hyperref[fig4]{4(d)} present the calculated energy-dependent bimagnon intensity along the same high-symmetry directions as Figs. \hyperref[fig4]{4(a)} and \hyperref[fig4]{4(b)}. Notably, the calculated bimagnon excitations exhibit a pronounced peak at $E \approx 6$ meV at $\Gamma$ [blue curve in Fig. \hyperref[fig4]{4(f)}], closely matching the observed excitation peak at the same energy [orange dashed curve in Fig. \hyperref[fig4]{4(f)}]. This suggests that the twofold-symmetric, $E \approx 6\pm1$ meV mode [Fig. \hyperref[fig3]{3(e),(j)}] originates from the density maximum of bimagnon excitations.

To further assess the role of bimagnons, we take the bimagnon peak at $E \approx 6\pm1$ meV as an intensity reference and compare the energy-dependent bimagnon intensity at $\Gamma$ [blue curve in Fig. \hyperref[fig4]{4(f)}] with the experimental excitation spectrum. By scaling the calculated $E \approx 6$ meV peak to match the observed excitation [orange dashed curve in Fig. \hyperref[fig4]{4(f)}], we find that bimagnon contributions account for only a small fraction of the broad excitation continuum around $\Gamma$. The remaining spectral weight forms a broad asymmetric peak [red dashed curve in Fig. \hyperref[fig4]{4(f)}], suggesting that the dominant part of the excitation continuum extends beyond conventional magnon excitations and may be attributed to fractionalized spin excitations. This corroborates the symmetry analysis concerning the nature of the magnetic excitations. 
Furthermore, the coexistence of these fractionalized excitations with the excitation gap below $T_N$ implies an interplay between low-energy spin waves and possible fractionalized spin states.

In summary, our results on partially detwinned sample provides crucial experimental insights into the complex nature of magnetic excitations in RuCl$_{3}$, revealing the necessity of additional interaction terms to fully capture its excitation spectrum and 
advancing the theoretical understanding of Kitaev materials. Furthermore, we demonstrate that biaxial anisotropic strain serves as a symmetry-breaking field that aligns magnetic domains across $T_N$, offering a 
broadly applicable detwinning approach for systems exhibiting $C_{2}$-symmetric magnetic order and magnetoelastic coupling. These findings establish a foundation for refined theoretical models of RuCl$_3$ and experimental methodologies for the study of Kitaev materials.

{\bf Acknowledgments}

This work is supported by National Key Projects for Research and Development of China with Grant No. 2021YFA1400400, the Scientific Research Innovation Capability Support Project for Young Faculty (Grant No. 2251300009), Beijing National Laboratory for Condensed Matter Physics (Grant No. 2024BNLCMPKF005), and the National Natural Science Foundation of China (Grants Nos. 11734002, 11922402, 12174029, 12374137, and 12434005). We gratefully acknowledge the Science and Technology Facilities Council (STFC) for access to neutron beam time at ISIS \cite{Lu2023LET}.

\end{document}